\documentclass[useAMS,usenatbib]{mn2e}

\usepackage[english]{babel}
\usepackage[ansinew]{inputenc}
\usepackage{amssymb,amsmath}
\usepackage{graphicx} 
\usepackage{natbib} 
\usepackage{color}
\usepackage{ulem}

\newcommand{\dx}{\Delta x}
\newcommand{\dy}{\Delta y}

\newcommand{\emath}{\end{displaymath}}

\newcommand{\beq}{\begin{equation}}
\newcommand{\eeq}{\end{equation}}
\newcommand{\bea}{\begin{eqnarray}}
\newcommand{\eea}{\end{eqnarray}}

\newcommand{\bfg}{\begin{figure}}
\newcommand{\efg}{\end{figure}}
\newcommand{\bitm}{\begin{itemize}}
\newcommand{\eitm}{\end{itemize}}
\newcommand{\bnum}{\begin{enumerate}}
\newcommand{\enum}{\end{enumerate}}
\newcommand{\btbl}{\begin{table}}
\newcommand{\etbl}{\end{table}}
\newcommand{\btbu}{\begin{tabular}}
\newcommand{\etbu}{\end{tabular}}

\title[Fast Shape Estimation for Galaxies and Stars]{Fast
  Shape Estimation for Galaxies and Stars}
\author[G. L. Li, B. Xin\& W. Cui (2012)]{Guoliang Li$^{1,2}$\thanks{E-mail:
guoliang@pmo.ac.cn}, Bo Xin$^{2}$ and Wei Cui$^{2}$\\
$^{1}$Purple Mountain Observatory, 2 West Beijing Road, Nanjing 210008, China\\
$^{2}$Department of Physics, Purdue University, 525 Northwestern Ave., West Lafayette, Indiana 47907, USA}

\begin{document}

\date{Accepted 20xx MM DD. Received 20xx MM DD.}

\pagerange{\pageref{firstpage}--\pageref{lastpage}} \pubyear{2010}

\maketitle

\label{firstpage}

\begin{abstract}
Model fitting is frequently used to determine the shape of galaxies
and the point spread function, for examples, in weak lensing analyses
or morphology studies aiming at probing the evolution of galaxies. 
However, the number of parameters in the model, as well as the number
of objects, are often so large as to limit the use of model fitting for future large surveys. In this article, we propose a set of algorithms to speed up the fitting process.
Our approach is divided into three distinctive steps: centroiding,
ellipticity measurement, and profile fitting. We demonstrate that we can
derive the position and ellipticity of an object analytically in the
first two steps and thus leave only a small number of parameters to be
derived through model fitting. 
The position, ellipticity, and shape parameters can then used in
constructing orthonomal basis functions such as s\'ersiclets for
better galaxy image reconstruction.
We assess the efficiency and accuracy of the algorithms with simulated
images. We have not
taken into account the deconvolution of the point spread function,
which most weak lensing analyses do. 

\end{abstract}


%%%%%%%%%%%%%%%%%%%%%%%%%%%%%%%%%%%%%%%%%%%%%%%%%%%%%%%%%%%%%%%%%%%%%%%%%%%

\section{Introduction}
\label{intro}

Quantifying shapes of galaxies and stars has long been one of the key
tasks in astronomical image analyses.
The shape of an
astronomical object is one of the a few direct observables.
A lot of useful information can be inferred from the shapes of galaxies and stars.
For example, galaxy mophology provides important knowledge on the
formation and evolution of galaxies (\citealp{morph1};~\citealp{morph2}).
As one of the most promising tools to probe dark energy and dark
matter in the universe, weak gravitational lensing relies on precision
measurements of the ellipticities of the background galaxies and the
point spread function (PSF). % (\citealp{bernstein2002}). 

The signal in weak lensing, however, is often very weak and noisy, due
to the large intrinsic dispersion in the shapes of background
galaxies. For ground-based observations, the shapes of galaxies are
further distorted by atmospheric turbulence and optical distortions,
which can be described by the PSF, and quantified using shapes of stars.
The first weak lensing algorithm was proposed by
\citet{ksb95} (hereafter KSB) and improved by~\citet{ksb97} and
\citet{ksb98}. Since then, a variety of algorithms were proposed and a
series of data analysis challenges have been carried out to improve
the precision and reduce systematic biases
(\citealp{step1};~\citealp{step2};~\citealp{great08},~\citealp{great10}
and the references therein). However, none of the algorithms satisfies
the requirements of future surveys such as the Dark Energy Survey (DES), 
the Canada-France-Hawaii-Telescope Legacy Survey (CFHTLS), 
and the Large Synoptic Survey Telescope (LSST).
Given the enormous amount of galaxies to be covered by these surveys,
one limiting factor of the existing algorithms is the accuracy and
efficiency with which the 
shapes and galaxies and stars are measured.

One of the existing powerful tools that has been studied in great detail is
the decompostion of images using basis functions, such as the
Gaussian-Laguerre expansion (\citealp{bernstein2002}),
or Shapelets
(\citealp{shapelet1};~\citealp{shapelet2};~\citealp{polarshapelet})
or s\'erscilets (\citealp{2009MNRAS.396.1211N};~\citealp{sersiclets}).
However, there are limitations to these methods,
because that the zeroth order image is often a poor match to real
galaxy profiles.  
While theoretically
any intensity profile can be written as a weighted sum of any complete
set of basis functions, in reality, 
the more the zeroth order resembles the real profile, the fewer basis
functions are needed in the decompostion(\citealp{polarshapelet};~\citealp{2010AJ....140..870B}).
As a first
step, it is necessary to accurately and efficiently quantify the {\it
  observed} shape of galaxies based on noisy images, which can then be
used as input in the construction of the basis functions.

In this article, we present a set
of efficient algorithms specifically for the task of parameter
estimations for observed galaxies, including center position,
ellipticity, and shape parameters such as size and steepness. 
We then run numerical tests to demonstrate the effectiveness of these
algorithms, and how the basis functions
could help further improve the image reconstrcution.
Our algorithms, developed with their weak lensing applications in mind,
  could also be used for others purposes such as to measure the
  approximate morphological parameters for large population of
  galaxies(e.g.,\citealp{2009MNRAS.393.1531G};~\citealp{2012MNRAS.423.3486W}). 
We leave PSF deconvolution to later work.

Usually, no less
than six parameters are needed to model the light distribution of an
object, including the centroid position $(x,y)$, the ellipticity
$(g_1,g_2)$, the normalization $I_0$, and the profile
parameter(s). High-dimensional parameter search is very time
consuming, especially when the number of objects is large, and tends to be trapped at local minima. Instead of brute-force fitting, we propose to derive the centroid position, ellipticity, and normalization of a light distribution numerically and thus reduce the number of parameters that need to be determined through fitting. Similar effort has been previously undertaken.
\citet{Miller2007} and \citet {Kitching2008} proposed a fast-fitting
algorithm in Fourier space, which also takes into account the effects
of PSF. Here, we will only focus on how to
reproduce the observed images. 

Once the observed shape of galaxies (and stars for PSF determination)
is described by smooth model(s), one may proceed to use the invariants
equation to arrive at the intrinsic shape of galaxies
(\citealp{F&K};~\citealp{deimos}), or to do the PSF deconvolution using basis
  function transformations 
(\citealp{polarshapelet};~\citealp{2009A&A...493..727M};~\citealp{2010AJ....140..870B}).
  In practice, the simple models used
to describe the light profile of stars and galaxies often lead biases
in the measurements
(\citealp{2010MNRAS.404..458V};~\citealp{2010A&A...510A..75M}). The
situation can be improved by adopting more sophisticated spatial
models or a set of basis functions. 
Our algorithms will be useful in providing input to the construction of basis functions (\citealp{licui2012}).
 
This article is organised as follows. We describe the algorithms for position and ellipticity determination in Sections~\ref{sec:center} and \ref{sec:g}, and numerical tests in Section~\ref{sec:tests}. We conclude by discussing
the limitation and prospects of our algorithm in Section~\ref{sec:conclusions}.

\section{Centroiding}
\label{sec:center}
For simplicity, we assume a Gaussian intensity profile and choose the origin of the coordinate system to be at the estimated center of the profile. With the actual center at ($\Delta x, \Delta y$), the intensity profile is given by
\beq
I(x,y) = I_0 e^{-\frac{(x-\Delta x)^2+(y-\Delta y)^2}{2\sigma^2}} +  \epsilon,
\eeq
where $\epsilon$ is the noise term. Now, with a Gaussian weight function, we define
\begin{align}
N & =  \int_{-\infty}^{\infty} I(x,y)
e^{-\frac{x^2+y^2}{2\sigma_w^2}}
dx dy,
\\
<x> & =  \int_{-\infty}^{\infty} I(x,y)
e^{-\frac{x^2+y^2}{2\sigma_w^2}}
x dx dy,
\\
<y> & =  \int_{-\infty}^{\infty} I(x,y)
e^{-\frac{x^2+y^2}{2\sigma_w^2}}
y dx dy.
\end{align}
Neglecting the noise terms, we have
\begin{align}
\frac{<x>}{N} & = k\dx,\\
\frac{<y>}{N} & = k\dy,
\end{align}
where $k=\sigma_w^2/(\sigma^2+\sigma_w^2)$. 

When the intensity profile has the same shape as the weight function, $\sigma=\sigma_w$,
we have $k=0.5$. If the former is extremely compact (i.e., $\sigma \approx 0$), we have
$k=1$. In general, the value of $k$ is determined iteratively.
For a true coefficient $0.5<k_{true}<1$, the choice of $k=0.75$ gives
$|\Delta_{i+1}|<\frac{1}{3}|\Delta_i|$, where $\Delta_i$ is the
distance between the centers of the intensity profile and the weight
function in the $i$th iteration. So, the calculation converges very
quickly. In typical cases, convergence is achieved in less than five
iterations, and the result can be further
  improved using an elliptical weight function with ellipticity
  obtained using the algorithm described in the next section.

\section{Ellipticity Measurement}
\label{sec:g}

Our approach is based on the KSB algorithm for shear measurement~(\citealp[hereafter KSB95]{ksb95};~\citealp{ksb97};~\citealp{ksb98}). Because our goal here is only to estimate the ellipticity of an object, there is no convolution or deconvolution involved. This means that we are only concerned with shear polarizability $P^{SH}$.

The KSB algorithm is perturbative in nature and is applicable only when the ellipticity is small. To adapt it for general ellipticity measurement, we have adopted an iterative approach. At each step, we evaluate
\beq
\Delta g = g-g_w,
\eeq
where $g$ and $g_w$ are the ellipticities at the current and previous steps, respectively. The difference is then added to $g_w$ to obtain a better estimate of $g$. The iteration continues until the difference is less than $10^{-4}$.

In contrast to the KSB algorithm, we use an elliptical weight function here, with an ellipticity also of $g_w$. It can be shown (see Appendix~\ref{app:ksbew}) that the shear polarizability tensor with elliptical weight function can still be written in the form
\beq
P^{SH}_{\alpha\beta}  = X^{SH}_{\alpha\beta} - e_\alpha e^{SH}_\beta.\label{eq:psh},
\eeq
with the new $X^{SH}_{\alpha\beta}$ and $e^{SH}_\beta$ given in Eqs.~(\ref{eq:xshew}) and Eq.~(\ref{eq:eshew}) in Appendix~\ref{app:ksbew}, respectively.

How do we estimate $\Delta g$ from $g_w$? As shown in Appendix~\ref{app:deltag}, with additional terms to $X^{SH}$ and $e^{SH}$, which we denote as $\Delta(X^{SH})$ and $\Delta(e^{SH})$, we have
\beq
\Delta g = g - g_w = (1-g^2_w)(P^{SH})^{-1}({e}^{\rm obs}-{e}_w), \label{eq:newg}
\eeq
where $e^{\rm obs}$ is the observed ellipticity which is calculated by using
 our elliptical weight function, and ${e}_w$ is the ellipticity corresponding to $g_w$.
 Eq.~(\ref{eq:newg}) makes it possible to measure the ellipticity of an object by iterating on the ellipticity of the weight function until it matches that of the object.

\citet{bernstein2002} proposed a similar idea but implemented
differently. They adopted a round weight function and try to shear the
image back to be round with $-g_w$. In that case, some knowledge of
the multipoles of the intensity profile and their derivatives (which
are not easy to compute) are needed.  \citet{hirata03} also
  calculated the centroid and the ellipticity using an adptive
  elliptical weight function. Our tests show that their algorithm
  takes much more iterations to converge than ours especially for ellipticity because
of the higher order corrections taken into account in our algorithm.

\section{Numerical Tests} 
\label{sec:tests}
\subsection{Galaxy Images}
\label{sec:simulation}
To test the methods, we have created a set of galaxy images based on the S\'ersic profile,
\beq
I(r) = I_0 e^{-b(\frac{r}{r_e})^\frac{1}{n}}, \label{eq:ser}
\eeq
where $b$ is chosen such that $r_e$ is the half-light radius. Therefore, only three independent parameters are required to specify the profile. For a given galaxy, a total of seven parameters are needed, including its centroid position (2 parameters) and ellipticity (2 parameters). The S\'ersic profile is one of the most popular functional forms that are used to model the light distribution of galaxies~\citep[e.g.,][]{Mac2003}.

For the simulation, we let $r_e$ vary between 2 and 15 pixels and $n$ vary between 0.5 and 5.
The ellipticity $g$ is limited to be no greater than 0.75. For small galaxies, $2<r_e<10$, we further require $n<0.5(r_e-2)+1$ to take into account PSF effects, which tend to flatten the inner profile. We also let the centroid position vary, with respect to the origin, in the range of $\dx,\dy\in(-20,20)$. Gaussian noise is added to maintain a constant signal-to-noise ratio ($=1000$) across an image. A total of 500 512 $\times$ 512 pixels images were produced for the tests. 

For each galaxy, we apply the methods described in Secs.~\ref{sec:center} and \ref{sec:g} to derive four of the seven parameters. Furthermore, the normalization factor may be derived from $\chi^2$ minimization,
\beq
\chi^2 = \sum_{i=1}^{N_{\rm pixel}}\left(\frac{I_i^{\rm
      pred}-I_i}{\sigma_i}\right)^2 \label{eq:chi2}.
\eeq
For the general case of $I_i^{\rm pred} = I_0 f_i^{\rm pred}\label{eq:II0}$, the minimization
\beq
\frac{\partial \chi^2}{\partial I_0} = 0 \label{eq:partialI0}
\eeq
leads to
\beq
I_0 = \frac{ \sum_{i=1}^{N_{\rm pixel}}I_i  f_i^{\rm pred}/\sigma^2_i }
{ \sum_{i=1}^{N_{\rm pixel}}  (f_i^{\rm pred})^2/\sigma^2_i }.
\eeq
Therefore, there are only two independent parameters ($n$ and $r_e$ for the S\'ersic profile) that need to be derived from model fitting. This greatly reduces the computation time. We used the {\it MINUIT} minimization program~\citep{minuit} in our implementation.

As a starting point, we estimate the half-light radius ($r_{50}$) by locating the
brightest pixel in an image and summing up the values of neighboring
pixels going outward, until the flux reaches half of the total flux. A
square region with sides six times $r_{50}$ is then cut out from the 512 $\times$ 512 grid. Subsequent image processing is carried out on the sub-image. For each galaxy, the centroid position is initially taken as the brightest pixel in the image.We then go through the iterative processes, as described in Secs.~\ref{sec:center} and \ref{sec:g}, to find the best centroid position and ellipticity. The starting values of $g_1$ and $g_2$ are estimated by the original KSB method. Iteration stops when the fractional change between two consecutive steps is less than $10^{-4}$ or the number of iterations reaches 20. 

\begin{figure}
\includegraphics[width=0.45\textwidth]{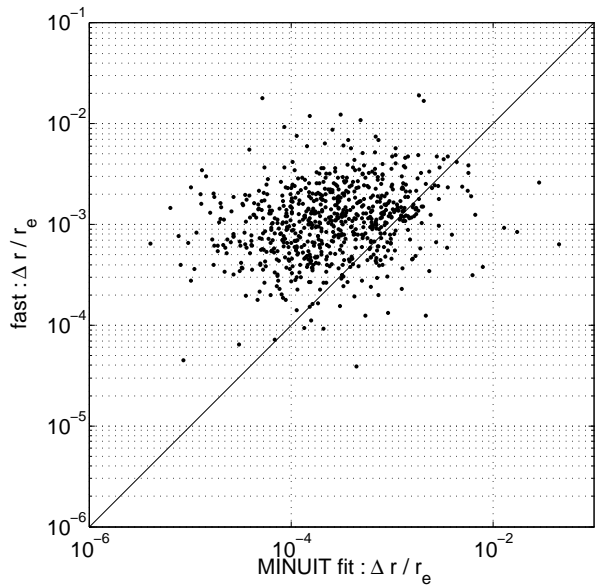}
\caption{Accuracy of centroiding. This shows a comparison between our
  approach (fast) and the brute-force method, based the S\'ersic profile. }
\label{fig:center}
\end{figure}

\begin{figure}
\includegraphics[width=0.45\textwidth]{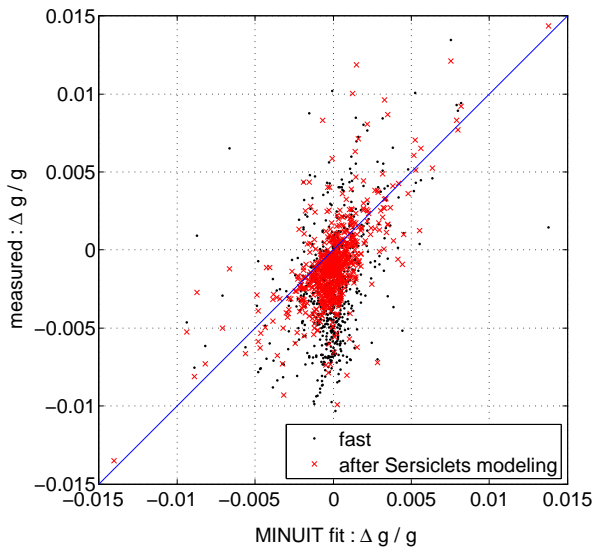}
\caption{Accuracy of ellipticity measurements. This shows a comparison
  between our approach (fast) and the brute-force method, based the S\'ersic
  profile. The dots are obtained using ellipticity algorithms
  decribed in Sec.~\ref{sec:g}, and crosses are results after using
  the nine lowest order s\'ersiclets basis functions constructed on a
  larger pixel grid. The side length of the larger pixel grid here is
  chosen to be ten times the half-light radius.}
\label{fig:g}
\end{figure}

\begin{figure}
\includegraphics[width=0.45\textwidth]{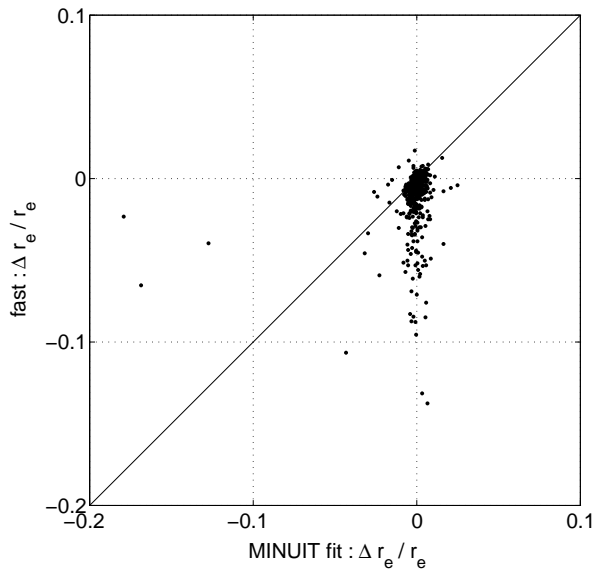}
\caption{Accuracy of half-light radius determination. This shows a
  comparison between our approach (fast) and the brute-force method, based the S\'ersic profile. }
\label{fig:re}
\end{figure}

\begin{figure}
\includegraphics[width=0.45\textwidth]{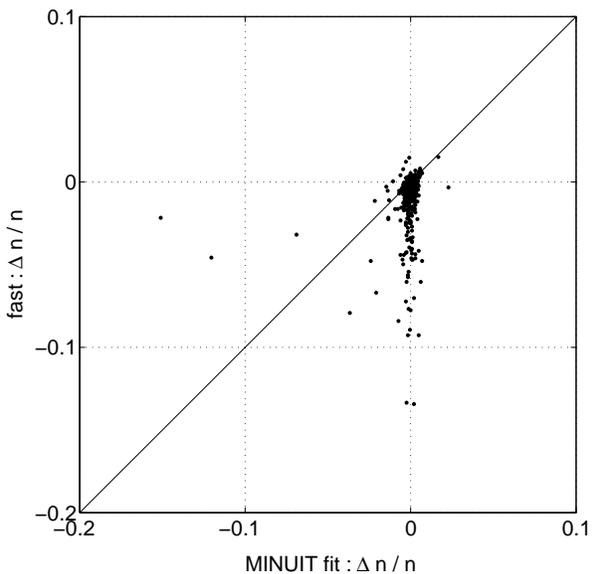}
\caption{Accuracy of S\'ersic index estimation. This shows a
  comparison between our approach (fast) and the brute-force method,
  based the S\'ersic profile.} 
\label{fig:n}
\end{figure}

To assess the accuracy and efficiency of our methods, we compare the results to those derived from the brute-force approach (i.e., 7-parameter fitting). Fig.~\ref{fig:center} show a comparison of centroiding, $\Delta r/r_e$, where $\Delta r=\sqrt{(\dx^{\rm est}-\dx)^2+(\dy^{\rm est}-\dy)^2}$ and ($\dx$,$\dy$) is the real center. Figs.~\ref{fig:g} - \ref{fig:n} show comparisons on $\Delta g/g$, $\Delta r_e/r_e$, and $\Delta n/n$, where $\Delta g=g^{\rm est}-g$, $\Delta r_e = r_e^{\rm est} - r_e$, and 
$\Delta n=n^{\rm est}-n$. The superscript, ``est'' indicates the corresponding
 values derived from our algorithms and brute-force fitting. 
The accuracy of our methods is slightly worse than that of brute-force
fitting. This can partially be attributed to the weight function used
in our case (but not in brute-force fitting), as the outer region may
play a non-negligible role. For more than 98\% of the cases, $\Delta
r/r_e$ from our method ranges between 0.01\% and 1\%. 

\begin{figure}
\includegraphics[width=0.45\textwidth]{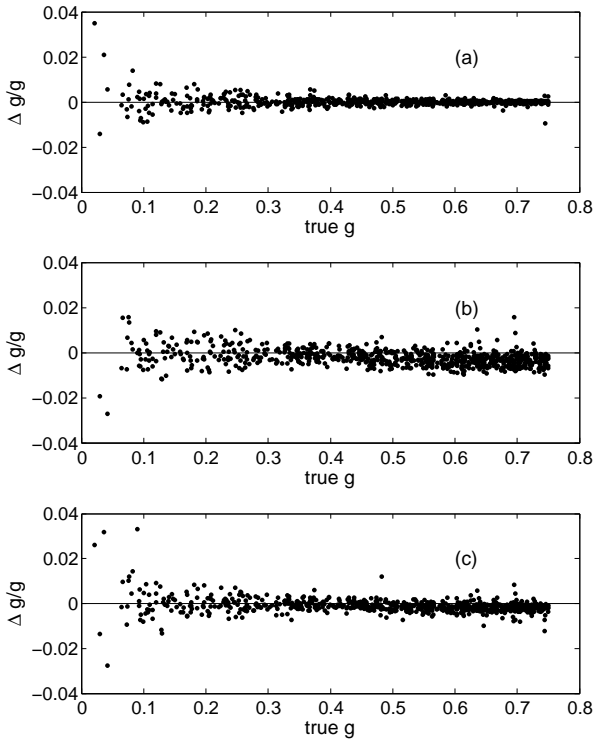}
\caption{The dependence of ellipticity estimation accuracy on the size
of ellipticity for (a) brute-force fitting, (b) our algorithms,
and (c) our algorithms combined with s\'ersiclets modelling.} 
\label{fig:g4}
\end{figure}

Our estimation of the ellipticity deviates from
the true ellipticity by less than 1\% in most cases.
Like any
moments-based ellipticity estimation algorithms, our estimation of $g$
is biased when $g$ is large, due to the limited size of
the image stamps used (side length 6$r_{50}$), resulting in moments
calculations not converging completely. 
However, this bias is significantly reduced if we use our measured
shape parameters to construct s\'ersiclets basis functions. The
reproduced image is extroplated to a bigger size (side length
10$r_{50}$) so that subsequent moments calculations
are not effected by the cut-off boundary. Then we run
our algorithms again on the s\'ersiclets model.\footnote{Such
  operations should be paid more attention to when the
  ellipticities of the outter and inner regions differ significantly.}
The crosses in Fig.~\ref{fig:g} shows the $\Delta g/g$ estimated using
the s\'ersiclets model with the nine lowest order basis functions
build on a larger pixel grid.
If we continue this process, each time we construct a new s\'ersiclets
model, the data points in Fig.~\ref{fig:g} continue to move toward the
diagonal line until it reaches an unbiased state.
Fig.~\ref{fig:g4} shows how $\Delta g/g$ varies with the true
ellipticity for brute-force fitting, our algorithms, and our algorithms
combined with s\'ersiclets modelling.
The effectiveness of our approach in removing the bias at large
ellipticity is clear. 
We also note that the wider spread of points in Fig.~\ref{fig:g4} (c)
than in (a) is because that the model used in brute-force
fitting is the same as what we used to simulate the images,
in which case the brute-force fitting is superior over any other
methods in terms of accuracy. But for real galaxies
that are typically more complex than a simple S\'ersic model (e.g., with
bulge, bar, spiral arms, etc.), the
basis functions will be more advantageous (e.g., \citealp{polarshapelet};~\citealp{sersiclets}).

Because the centroid position and ellipticity are both fixed in the
subsequent two-parameter fitting, the accuracy on $r_e$ and $n$ is
worse than that of the brute-force approach, but is still well within
10\% in most cases, as shown in Fig.~\ref{fig:re} and
  Fig.~\ref{fig:n}. Most of the the outliers have small $r_e$ and big
  $n$. Their images are dominated just by several central pixels. This
leads to relatively large uncertainty on the centrioding and then biases $r_e$
and $n$ bacause the strong degeneracy between these two parameters.

\begin{figure}
\includegraphics[width=0.45\textwidth]{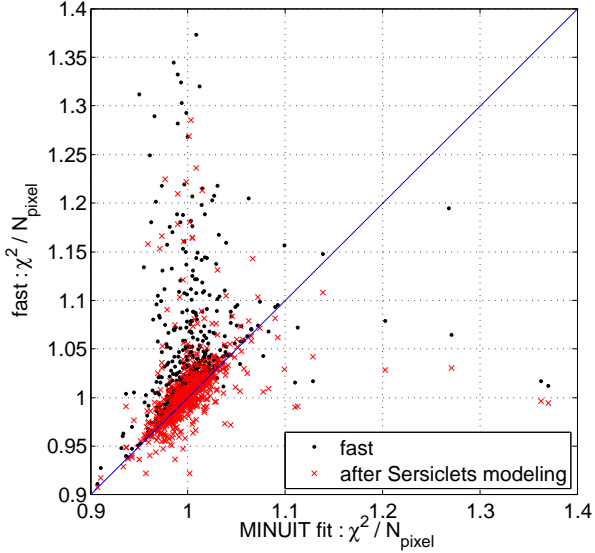}
\caption{A comparison of the reduced $\chi^2$ values between our
  method (fast) and the brute-force method, based on the S\'ersic profile.
The dots are obtained using algorithms described in Secs.~\ref{sec:center} and
\ref{sec:g}, and crosses are results after using the nine lowest order
s\'ersiclets basis functions.
}
\label{fig:chi2}
\end{figure}

Going beyond individual parameters, we assess the quality of the fit
to the overall light profile of a galaxy. Fig.~\ref{fig:chi2} shows a
comparison of the reduced $\chi^2$ values. When $N_{\rm pixel}$ is
large, the ratio $\chi^2/N_{\rm pixel}$ is expected to be around unity
if the estimation is unbiased. We found that the fit becomes
progressively worse, as the S\'ersic profile steepens (i.e., large
$n$) and the size decreases. When the half-light radius approaches the
size of a pixel, small error in centroiding may contribute
significantly to the $\chi^2$ value. 
As one would have expected from the results in Fig.~\ref{fig:g},
constructing s\'ersiclets model using the estimated parameters
reduces $\chi^2$, as shown by the crosses in Fig.~\ref{fig:chi2}.

As for efficiency, on a single-core 2.2 GHz Intel CPU, it takes about 0.1 seconds to carry out centroiding and ellipticity measurement on one galaxy image and about 0.5 seconds to derive the remaining two parameters in the S\'ersic model from profile fitting. In comparison, it takes about 6 seconds to process one galaxy image in the brute-force approach. 

\subsection{Star Images}
\label{sec:sigma}
We carried out similar tests with simulated star images, which are relevant to PSF modelling. Here, instead of using fitting programs like MINUIT, we developed faster numerical methods also to derive shape parameters from the moments of {\it observed} light profiles. Two kind of PSF profiles are considered here.

\subsubsection{Gaussian profile}
\label{sec:gau}

For the Gaussian profile,
\beq
I(r) = I_0e^{-\frac{r^2}{2\sigma^2}}, \label{eq:gau}
\eeq
the zeroth and second order moments are
\begin{align}
r_0 & = \int_0^\infty I(r)W(r) 2\pi r dr, \label{eq:gaur0}\\
r_2 & = \int_0^\infty I(r)W(r) 2\pi r^3 dr, \label{eq:gaur2}
\end{align}
where the weight function is
\beq
W(r) = e^{-\frac{r^2}{2\sigma_w^2}}.
\eeq

If we take $\sigma_w = \sigma$, we have
\begin{align}
r_0 & = \pi\sigma^2 I_0,\\
r_2 & =\pi\sigma^4 I_0.
\end{align}
Therefore,
\begin{align}
\sigma & = \sqrt{\frac{r_2}{r_0}}, \label{eq:gausigiter}\\
I_0 & = \frac{r_0}{\pi\sigma^2}. \label{eq:gauI0iter}
\end{align}

For the simulation, the half-light radius ($r_{50}$) is allowed to vary from 1.5 to 2.5 pixels. The ellipticity $g$ is limited to no greater than 0.25. We also let the centroid position vary, with respect to the origin, in the range of $\dx,\dy\in(-20,20)$. Gaussian noise is added to maintain a constant signal-to-noise ratio ($=300$) across an image. A total of 500 512 $\times$ 512 images are produced for the tests. 

As for galaxies, we carry out centroiding and ellipticity determination with the methods described in Secs.~\ref{sec:center} and \ref{sec:g}.  We then use Eqs.~(\ref{eq:gausigiter}) and (\ref{eq:gauI0iter}) to derive $\sigma$ and $I_0$ from the zeroth- and second-order moments. Setting $\sigma_w = \sigma$, we recalculate the moments. The iteration continues until the fractional change in $\sigma$ between two consecutive steps is less than $10^{-4}$ or the number of iterations reaches 20. 

\begin{figure}
\includegraphics[width=0.45\textwidth]{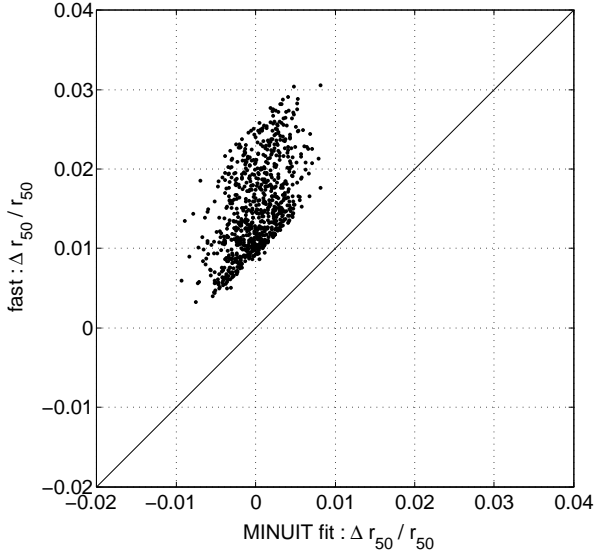}
\caption{Half-light radius. This shows a comparison between our
  approach (fast) and the brute-force method, based the Gaussian profile.}
\label{fig:gaur50}
\end{figure}

\begin{figure}
\includegraphics[width=0.45\textwidth]{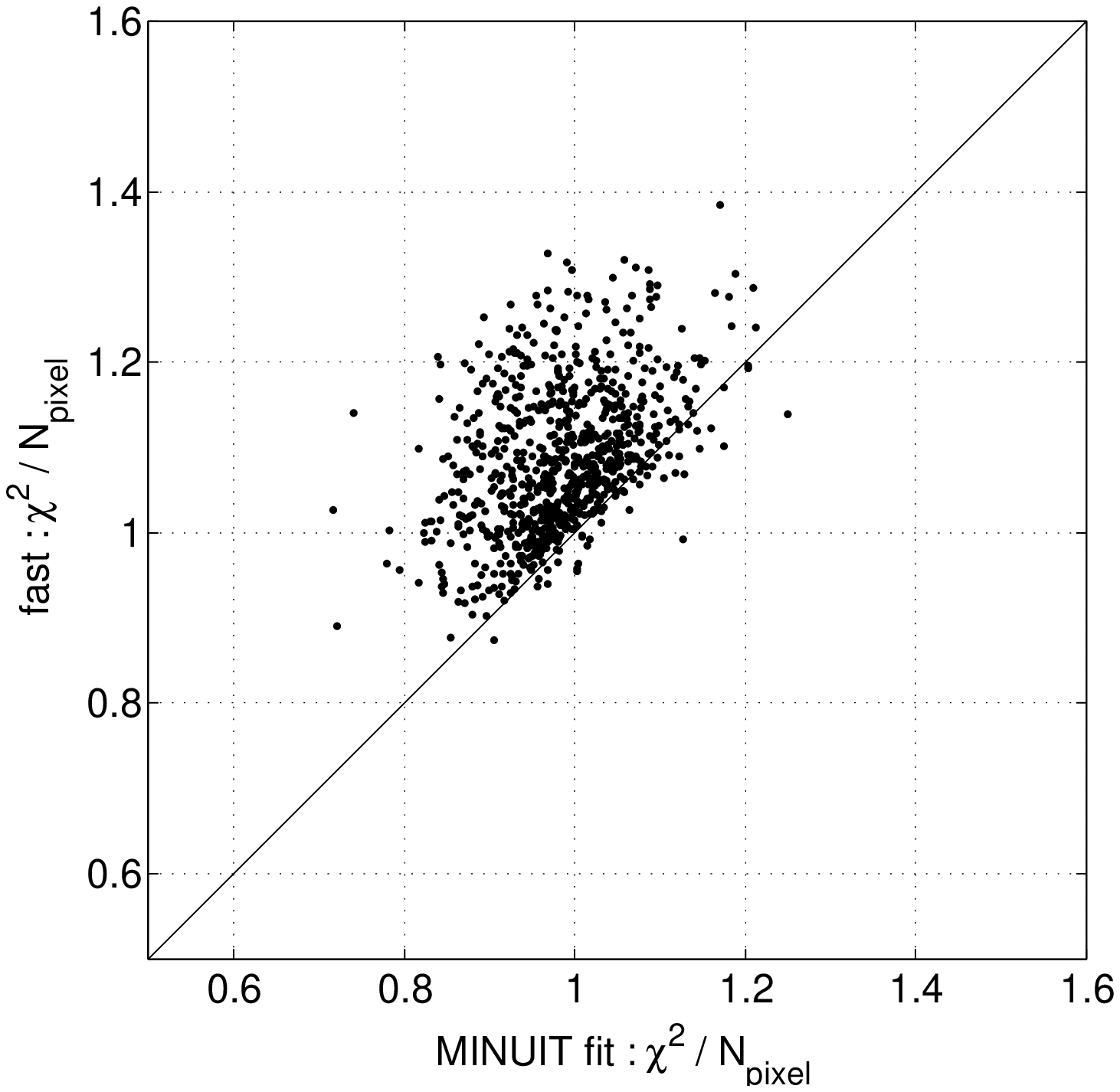}
\caption{A comparison of the reduced $\chi^2$ values between our
  method (fast) and the brute-force method, based on the Gaussian profile. }
\label{fig:gauchi2}
\end{figure}

Fig.~\ref{fig:gaur50} shows a comparison between our approach and brute-force fitting in determining the half-light radius. In practice, the integrations of Eqs.~(\ref{eq:gaur0}) and (\ref{eq:gaur2}) were done numerically. And this will bias the result because of pixelation, which has a prominent effect on small images. However, deviation is still within a few percent. The bias in the overall size can be roughly estimated as $\delta r_{50}^2/<r_{50}^2>~\lesssim0.001$. This value is acceptable for weak lensing statistics (\citealp{Paulin2009}). The overall quality of the shape estimation is shown in Fig.~\ref{fig:gauchi2}.

\subsubsection{Moffat profile}

For the Moffat profile
\beq
I(r) = I_0\left(1+\left( \frac{r}{r_d}
  \right)^2\right)^{-\beta}, \label{eq:mof}
\eeq
in addition to $r_0$ and $r_2$, we define
\beq
r_4=  \int_0^\infty I(r) W(r)  2\pi r^5 dr,
\eeq
where the weight function $W(r)$ is also of the Moffat shape,
\beq
W(r) = \left(1+\left( \frac{r}{r_{dw}} \right)^2\right)^{-3}.
\eeq

If we take $r_{dw} = r_d$, we have
\begin{align}
r_0 & = I_0 \frac{\pi r_d^2}{\beta+2}, \\
r_2 & = I_0\frac{\pi r_d^4}{(\beta+1)(\beta+2)}, \\
r_4 & = I_0 \frac{2\pi r_d^6}{\beta(\beta+1) (\beta+2)}.
\end{align}
Therefore,
\begin{align}
\beta & = \frac{1}{\frac{\displaystyle r_4r_0}{\displaystyle 2r_2^2}
  -1}, \label{eq:mofbetaiter}\\
r_d & = \sqrt{\frac{r_2}{r_0}(\beta+1)}, \label{eq:mofrditer}\\
I_0 & = \frac{r_0(\beta+2)}{\pi r_d^2}. \label{eq:mofI0iter}
\end{align}

\begin{figure}
\includegraphics[width=0.45\textwidth]{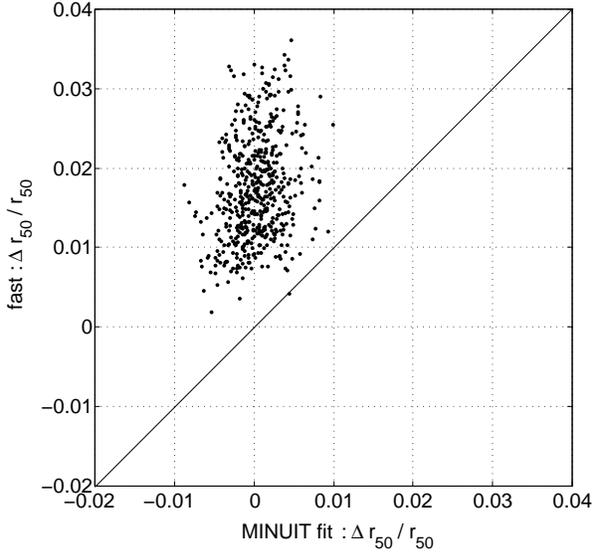}
\caption{A comparison of the determination of half-light radius
  between our method (fast) and the brute-force method. 
This is based on the Moffat profile.}
\label{fig:mofr50}
\end{figure}

\begin{figure}
\includegraphics[width=0.45\textwidth]{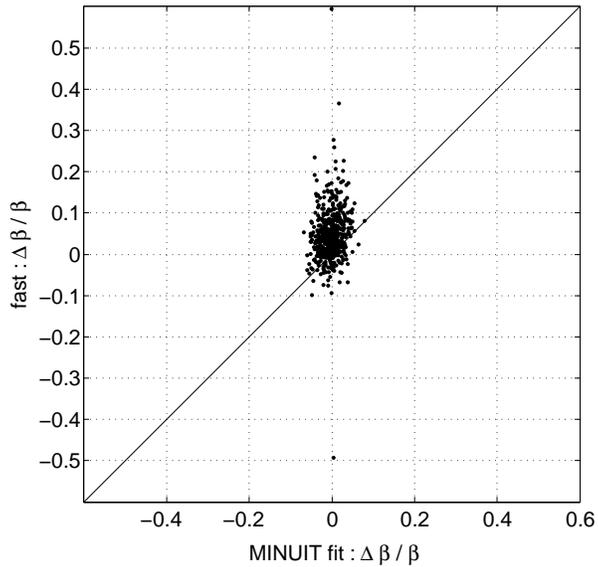}
\caption{A comparison of the determination of Moffat power-law index
  between our method (fast) and the brute-force method.}
\label{fig:mofbeta}
\end{figure}

\begin{figure}
\includegraphics[width=0.45\textwidth]{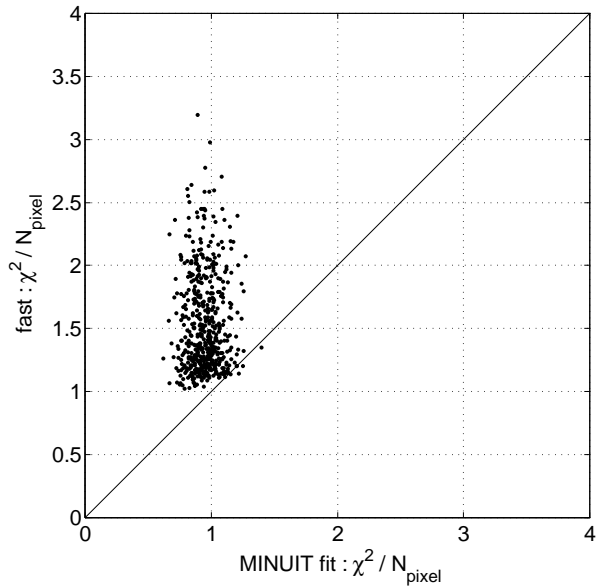}
\caption{A comparison of the reduced $\chi^2$ values between our
  method (fast) and the brute-force method, based on the Moffat profile}
\label{fig:mofchi2}
\end{figure}

As for the Gaussian profile, we have made 500 512 $\times$ 512 images. In the simulation, $\beta$ varies between 3 and 5. Similarly, we compute the centroid position, ellipticity, and $r_d$ from the moments by iterating on $r_{dw}$. The results are shown in Figs.~(\ref{fig:mofr50}) $-$ (\ref{fig:mofchi2}).
 
Again, the biases in $r_{50}$ and $\beta$ are attributable to pixelation effects
when we do the numerically integration of moments. The half-light radius $r_{50}$ is generally accurate to about 3-4\% and the power-law index $\beta$ to less than 20\%. As before, the bias in the overall size is roughly $\delta r_{50}^2/<r_{50}^2>~\lesssim0.001$.
Overall, the shape estimation is worse than in the case of Gaussian
profiles (comparing Fig.~\ref{fig:mofchi2} with
Fig.~\ref{fig:gauchi2}). We found that the points with $\chi^2 >1.5 $
are dominated by realizations with $r_d<2$. This is because that the
pixelation effect bias the moments two much. As same reason for
S\'ersic model, the Moffat profile also suffers from $r_d$-$\beta$
degeneracy. Therefore the problem here is more tough than that in
Gaussian profile. In practice, for the data taken under similar
weather conditions, we suggest to use several bright stars to derive
$r_d$ and $\beta$ by using fitting process and then fix $\beta$ at the
average value for other (fainter) stars. This could be a way out to
break the degeneracy and to decrease the pixelation effect.

Based on the shape estimation algorithms described and tested above
for the Gaussian and Moffat profiles, we have constructed two set of
basis functions which are named gaussianlets and moffatlets,
whose zeroth order profiles are Gaussian and Moffat,
respectively (\citealp{licui2012}). 
The corresponding algorithms have been tested in the
GRavitational lEnsing Accuracy Testing 2010 (GREAT10) 
Star Challenge (\citealp{great10}).
Our gaussianlets worked very well with accuracy of
$\sigma(e) \approx \sigma(R^2)/R^2 \approx 0.1\%$, where $e$ is the
ellipticity and $R$ is the size (\citealp{great10star}). But the moffatlets
didn't do very well especially when the PSF size is small.

\section{Discussion}
\label{sec:conclusions}

In this work, we have adapted the KSB algorithm for ellipticity determination. The new algorithm allows the use of elliptical weight functions and is applicable to highly elliptical shapes. This, combined with centroiding, makes it possible to eliminate four of the parameters from brute-force fitting. Consequently, less time is needed in searching for the remaining (shape) parameters. Overall, the efficiency is improved roughly by an order of magnitude.

We have tested our algorithms with simulated images that are
representative of stars and galaxies. In general, our centroiding
algorithm is accurate to well within 1\%, although it is slightly
worse than the brute-force fitting. For galaxy images,
which are made with the S\'ersic profile, our approach is relatively
worse than brute-force fitting in determining the half-light radius
and the S\'ersic index. This is mainly due to the fact that the two
parameters are tightly coupled, and then the fitting results will be
biased by any small offset in centroiding. Overall, we can find very good fit to
simulated galaxy profiles with our algorithms (as measured by the
reduced $\chi^2$). The accuracy of parameter estimations is seen to
be further improved when the algorithms are run on 
s\'ersiclets models with nine lowest order
basis functions which are constructed with the estimated parameters as input.
Given that galaxies are not perfectly S\'ersic-like, our algorithms,
combined with s\'ersiclets modelling, has unique advantages over other
shape estimation techniques.

For star images, we have adopted profiles that are often used for PSF modelling, Gaussian and Moffat. In this case, we show that we can also derive the shape parameters directly from the moments of the light distribution, which improve the efficiency even further.  
However, for small images, pixelation leads to significant
biases. Fortunately, the bias in the overall size is $\delta
r_{50}^2/<r_{50}^2>~\lesssim0.001$, which is still acceptable for weak
lensing statistics. We participated in the GREAT10 star challenge. The
results show that our methods performed quite well with Gaussian
profiles but not as satisfactorily with Moffat profiles, particularly
for data sets with stars of small radii. This tells us that the effects of pixelation and $r_d$-$\beta$ degeneracy must be taken into account for Moffat profile.

Like other algorithms, we also assume axisymmetric light distributions with constant ellipticity. In practice, this is known not to be the case for either galaxies or PSFs. In fact, the morphology of a galaxy or PSF can be quite complex.Therefore the use of simple spatial models leads to biases. The accuracy is expected to improve with more complicated spatial models. One possible option is to adopt appropriate basis functions, such as s\'ersiclets, moffatlets and gaussianlets, to reconstruct the light distribution. The approach described here shows its strong capability in providing input parameters for the basis functions. In spite of the deficiencies, our algorithms are very fast, which is important for future surveys, as the data volume is expected to increase drastically. 

\section*{Acknowledgements}
This work was supported in part by the U.S.\ Department of Energy through Grant DE-FG02-91ER40681. We are grateful to support from Purdue University.
GL is also supported by the one-hundred talents program of the Chinese
Academy of Sciences (CAS). We would like to thank the anonymous
referee for the constructive and clarifying comments.
\bibliographystyle{aa}

\def\physrep{Phys. Rep.}%
          % Astrophysical Journal, Supplement
\def\apjs{ApJS}%
          % Astrophysical Journal, Supplement
\def\apj{ApJ}%
          % Astrophysical Journal
\def\aj{AJ}%
          % Astronomical Journal
\def\aap{A\&A}%
          % Astronomy and Astrophysics
\def\aaps{A\&AS}%
          % Astronomy and Astrophysics Supplements
\def\mnras{MNRAS}%
          % Monthly Notices of the RAS
\def\araa{ARA\&A}%
          % Annual Review of Astron and Astrophys
\def\pasa{PASA}%
\def\nat{Nature}%
\bibliography{bibliography}

\onecolumn{}
\appendix

\section{The KSB algorithm with an elliptical weight function}
\label{app:ksbew}
Lets $W(R^2)$ be an elliptical weight function, which is assumed to be
converted by shear polarization $g_w$,
\beq
R^2=[(1-g_{w1})^2+g_{w2}^2]\theta^2_1+[(1+g_{w1})^2+g_{w2}^2]\theta^2_2-4g_{w2}\theta_1\theta_2. 
\eeq
Eq. (B5) in KSB95 now takes on the form
\begin{align}
z_{lmij} & =  \frac{\partial [W({\bmath \theta}) \theta_i \theta_j \theta_m]}
{\partial \theta_l} \nonumber \\
& = 
W(\delta_{il} \theta_j \theta_m + \delta_{jl} \theta_i \theta_m + 
\delta_{ml} \theta_i \theta_j) + W' \theta_i \theta_j \theta_m 
\frac{\partial (R^2)}{\partial \theta_l}.
\end{align}
Compared to the cases with circular weight function, the only change is that $2\theta_l$ is replaced with $ \frac{\partial (R^2)}{\partial  \theta_l}$ in the terms that contain $W'$, where the prime denotes
differentiation with respect to $R^2$,
i.e., 
\begin{align}
\theta_1 & \longrightarrow 
\frac{1}{2}\frac{\partial (R^2)}{\partial  \theta_1} =
 A\theta_1 -2g_{w2}\theta_2, \\
\theta_2 & \longrightarrow 
\frac{1}{2}\frac{\partial (R^2)}{\partial  \theta_2} =
 B\theta_2 -2g_{w2}\theta_1,
\end{align}
where $A=1+g^2_w-2g_{w1}$ and $B=1+g^2_w+2g_{w1}$.
$X^{SH}$ and $e^{SH}$ are now given by
\beq
X^{SH}  = 
\frac{1}{T} \int d^2 \theta f(\bmath{\theta}) 
\left(
\begin{array}{cc}
2W\theta^2+2W'(\theta_1^2-\theta_2^2)(A\theta_1^2-B\theta_2^2) & 
2W' (\theta_1^2-\theta_2^2)[(A+B)\theta_1\theta_2-2g_{w2}\theta^2] \\
4W'\theta_1\theta_2 (A\theta_1^2-B\theta_2^2) &
2W\theta^2+4W'\theta_1\theta_2
[(A+B)\theta_1\theta_2-2g_{s2}\theta^2]
\end{array}
\right),  \label{eq:xshew}
\eeq
and
\beq
e^{SH}  = 
2\left(
\begin{array}{c}
e_1 \\
e_2 
\end{array}
\right)
+
\frac{2}{T} 
\int d^2 \theta f(\bmath{\theta}) 
W'\theta^2\left(
\begin{array}{c}
A\theta_1^2-B\theta_2^2 \\
(A+B)\theta_1\theta_2-2g_{w2}\theta^2
\end{array}
\right).  \label{eq:eshew}
\eeq

\section{The iterative process for ellipticity determination}
\label{app:deltag}
The KSB method is applicable only when the ellipticity of an object is small, i.e., $\phi_{,ij}$ in Eq. (B7) in KSB95 is small, so that the Taylor expansion is appropriate. However, galaxies are often highly elliptical. This appendix shows a derivation of $\Delta g=g-g_w$, where $g$ is the intrinsic ellipticity of an object and $g_w$ is the estimated ellipticity.

Let ${\bmath \beta}$ be the coordinates in the virtual plane with ellipticity $g_w$. ${\bmath \beta}$ and ${\bmath \theta}$ are related by
\beq
{\bmath \beta} = C {\bmath \theta}.
\eeq
where
\beq
C =
\left(\begin{array}{cc}
1-\gamma_1-(g_{w1}\gamma_1+g_{w2}\gamma_2)&
  -\gamma_2+g_{w2}\gamma_1-g_{w1}\gamma_2\\
  -\gamma_2-g_{w2}\gamma_1+g_{w1}\gamma_2&
1+\gamma_1-(g_{w1}\gamma_1+g_{w2}\gamma_2)
\end{array}\right), \label{eq:C}
\eeq
and 
\beq
\gamma_1 = \frac{\Delta g_1}{1-g^2_w}, \quad 
\gamma_2 = \frac{\Delta g_2}{1-g^2_w},
\eeq

The surface brightness is conserved during the transformation defined
in Eq.~(\ref{eq:C}),
\beq
f_w({\bmath \beta}) = f({\bmath \theta}).
\eeq
Let $\psi=C^{-1}-1$, it follows that
\beq
f_w(\bmath{\beta}) = f(C^{-1}\bmath{\beta}) = f(\bmath{\beta}+\psi \bmath{\beta}).
\eeq
Following the original KSB formalism, 
\beq
Q^w_{ij} = Q_{ij} -  \psi_{lm} Z_{lmij},
\eeq
where $Z_{lmij}$ is defined in Eq. (B4) in KSB95. In their Eqs. (3.2) and (3.3), the difference
in ellipticity $e$ between the real image with reduced shear $g$ and
the virtual image with $g_w$ is easily related to $\delta Q_{ij} = Q_{ij} - Q^w_{ij}$.
We then find that, in addition to the terms defined in
Eqs.~(\ref{eq:xshew}) and (\ref{eq:eshew}) in Appendix~\ref{app:ksbew},
we now have the following extra terms on $X^{SH}$ and $e^{SH}$:
\begin{align}
\Delta(X^{SH}) & = 
\frac{1}{T}\int d^2\theta f(\bmath{\theta}) \left(
\begin{array}{cc}
-g_1 P_1 - g_2 P_2&
g_1 P_2 - g_2 P_1 \\
-g_1 P_3-g_2 P_4 &
g_1  P_4 - g_2 P_3
\end{array}
\right), \label{eq:dxsh}
\end{align}
and
\begin{align}
\Delta(e^{SH}) =
\frac{1}{T} \int d^2 \theta f(\bmath{\theta}) 
\left(
\begin{array}{c}
-g_1 P_5-g_2 P_6 \\
g_1 P_6 -g_2 P_5
\end{array}
\right), \label{eq:desh}
\end{align}
where
\begin{align}
P_1 &=
4W(\theta_1^2-\theta_2^2)+2W'(\theta_1^2-\theta_2^2)
(A\theta_1^2+B\theta_2^2-4g_2\theta_1\theta_2), \\
P_2 &=
4W\theta_1\theta_2+2W'(\theta_1^2-\theta_2^2)
[(A-B)\theta_1\theta_2+2g_2(\theta_1^2-\theta_2^2)], \\
P_3 &=
8W\theta_1\theta_2+4W'\theta_1\theta_2
(A\theta_1^2+B\theta_2^2-4g_2\theta_1\theta_2), \\
P_4 &=
-2W(\theta_1^2-\theta_2^2)+4W'\theta_1\theta_2
[(A-B)\theta_1\theta_2+2g_2(\theta_1^2-\theta_2^2)],\\
P_5 &=
4W\theta^2+2W'\theta^2
(A\theta_1^2+B\theta_2^2-4g_2\theta_1\theta_2), \\
P_6 &=
2W'\theta^2
[(A-B)\theta_1\theta_2+2g_2(\theta_1^2-\theta_2^2)].
\end{align}

\bsp

\label{lastpage}

\end{document}